\documentclass[10pt]{iopart}

\usepackage{amsmath,amsfonts}
\usepackage{amsgen}
\usepackage{amssymb}
\usepackage{algorithmic}
\usepackage{algorithm}
\usepackage{array}

\usepackage{textcomp}
\usepackage{stfloats}
\usepackage{url}
\usepackage{verbatim}
\usepackage{graphicx}
\usepackage{cite}
\usepackage{bm}
\usepackage{siunitx}

\usepackage{algorithm}
\usepackage{verbatim}
\usepackage{stfloats}
\usepackage{caption}
\usepackage{subcaption}
\usepackage{multirow}
\usepackage{multicol}
\usepackage{comment}
\usepackage{upgreek}
\usepackage{rotating}
\usepackage{color}
\usepackage{ragged2e}
\usepackage{colortbl}
\usepackage{hhline}

\usepackage{titlesec}

\begin{document}

\title[]{Optimized Bistable Vortex Memory Arrays for Superconducting In-Memory Matrix-Vector Multiplication}

\author{M. A. Karamuftuoglu, C. Song, B. Z. Ucpinar, S. Razmkhah and M. Pedram}

\address{Ming Hsieh Department of Electrical and Computer Engineering, University of Southern California, Los Angeles, CA 90089 USA\\}
\ead{karamuft@usc.edu}
\vspace{10pt}
\begin{indented}
\item[]March 2025
\end{indented}

\begin{abstract}
Building upon previously introduced Bistable Vortex Memory (BVM) as a novel, nonvolatile, high-density, and scalable superconductor memory technology, this work presents a methodology that uses BVM arrays to address challenges in data-driven algorithms and neural networks, specifically focusing on matrix-vector multiplication (MVM). The BVM approach introduces a novel superconductor-based methodology for in-memory arithmetic, achieving ultra-high-speed and energy-efficient computation by utilizing BVM arrays for in-memory computation. The design employs a tiled multiplier structure where BVM's inherent current summation capability is combined with Quantizer Buffer (QB) cells to convert the analog accumulated current into a variable number of digital Single Flux Quantum (SFQ) pulses. These pulses are then processed by T1 adder cells, which handle binary addition and carry propagation, thereby forming a complete functional multiplier unit. This paper thus presents an efficient MVM architecture that uses these BVM-based multipliers in a systolic array configuration to enable parallel computation. A key innovation is an optimized BVM array structure specifically tailored for multiplication applications, involving a restructuring of Sense Lines (SLs) with diagonal connections to reduce area and an adjusted input scheme to enhance computational efficiency compared to the general-purpose BVM array design. We demonstrate the efficacy of this approach with a 4-bit multiplier operating at 20 GHz with 50 ps latency and an MVM structure demonstrating operation at 20 GHz. Furthermore, we showcase how this multiplier design can be extended to support Multiply-Accumulate (MAC) operations. This work paves the way for power-efficient neural networks by enabling high-speed in-memory computation.
\end{abstract}

\vspace{2pc}
\maketitle
\ioptwocol

\section{Introduction}
Superconductor electronics (SCE) offer an emerging alternative to traditional computing systems, using ultrafast operating speeds and low power consumption \cite{holmes2021cryogenic}. Recent advances in SCE designs have further improved scalability and integration, driving the development of computing paradigms beyond traditional CMOS-based approaches \cite{razmkhahBook}. In particular, Rapid Single Flux Quantum (RSFQ) logic \cite{likharev1991a}, and Adiabatic Quantum Flux Parametron (AQFP) \cite{takeuchiAQFP2013} have been extensively explored for high-throughput and energy-efficient computation. However, the lack of compact memory, large cell sizes, and the fanout of one make it challenging for complex and large-scale operations \cite{razmkhah2024challenges}.

Current superconductor multiplier designs are implemented using digital logic gates and achieve high throughput by deep pipelines. However, as computational complexity increases, these designs face significant challenges. In particular, the input fanin needs large splitter trees. Deeply pipelined digital architectures require extensive clock tree distribution and path balancing to maintain synchronization, which adds to design complexity and resource overhead, and an increase in size results in higher clock skew and a drop in throughput \cite{Haolin_Multiplier_2021, Nagaoka_Multiplier_2021, Yamanashi_Multiplier_2024}. For example, in a 64-bit multiplier design, more than 80\% of the cells used are for splitter trees and DFFs balancing the paths \cite{Coldflux2023}.

Matrix-vector multiplication (MVM) is fundamental in numerous data-driven algorithms, neural networks, and signal processing applications \cite{McCanny_MVM_1983, Alam_Superconducting_MVM_2023}. Although conventional superconductor logic circuits can achieve high throughput due to a deeply pipelined structure, the large size of the cells results in a large latency. The long latency and lack of memory density result in huge costs when memory access or fetch cycles are involved. The von Neumann architecture with separate memory and logic regions further limits conventional approaches due to its memory access latency and inefficient data handling, constraining overall performance. Overcoming these challenges requires hardware designs that eliminate the separation of memory and data processing regions while ensuring high-speed and efficient computation with minimal power consumption.

In-memory computing (IMC) research now follows two converging but technologically distinct tracks \cite{zolfagharinejad2024brain, bao2022toward}. On the volatile side, the idea of moving computation into DRAM was flagged by Siegl et al. as an urgent remedy for the memory-bandwidth wall \cite{siegl2016data}. A concrete embodiment—row-level charge-sharing bitwise logic—was later demonstrated by Ambit in commodity DRAM devices \cite{seshadri2017ambit}. On the non-volatile side, memristor crossbars integrate storage and multiply-paccumulate in a single Ohmic network: Chua's theoretical memristor \cite{chua_memristor-missing_1971} and the TiO$_2$ experimental breakthrough \cite{strukov2008missing,williams2008we} seeded a rich device literature \cite{sun2019understanding, zidan2018future, lee2018demand}, culminating in analog crossbar tutorials that survey large-scale ReRAM/PCM/FeFET accelerators for neural and scientific workloads \cite{yao2020fully,han2020electric,jung2022crossbar}. DRAM-PIM offers byte-level precision atop a mature manufacturing stack but is bounded by bit-line current limits. In contrast, memristive IMC delivers massively parallel analog MACs and non-volatility at the expense of device variability and ADC/DAC overhead. 

In this work, we present a novel superconducting MVM design that utilizes bistable vortex memory (BVM) \cite{Karamuftuoglu_BVM_2025} and T1 \cite{bairamkulovT1} cells as fundamental components for the multiplication process. The superconducting BVM architecture introduced in this work inherits IMC's data-proximity advantage yet sidesteps both constraints, providing cryogenic-speed, digital-precision operations without peripheral conversion—thereby positioning itself as a complementary third pathway toward energy-proportional computing. The BVM enables the intermediate summation of partial products, ensuring proper signal amplitude scaling on the intermediate result for practical carry propagation in the next computational stage, while the T1 cells generate a signal via a cascaded adjustment mechanism, finalizing the multiplication result. Furthermore, we extend the multiplier design to support MAC operations. This approach effectively reduces complexity and latency while improving overall performance.

The proposed MVM design utilizes a systolic array configuration to enable parallel computation across processing elements (PEs). Each PE integrates a single-cycle 4-bit multiplier based on BVM, supporting fast and efficient MVM operations. Unlike designs incorporating exotic materials or ferromagnetic Josephson junctions (JJs), our approach relies on standard JJs, ensuring compatibility with conventional superconducting fabrication processes. By leveraging the inherent benefits of superconducting circuits, the architecture achieves ultra-high-speed computation with a throughput of 20 GHz and a latency of 200 ps, significantly outperforming traditional digital implementations. Furthermore, this work explores the scalability of the design, highlighting its potential for larger and more complex neural networks and signal processing applications.

The key contributions of this paper are as follows.
\begin{itemize}
\item Development of a superconducting multiplier that incorporates BVM and T1 cells.
\item Extending the multiplier design to support high-speed and scalable MAC operation.
\item Implementation of a high-performance MVM structure supporting parallel computations through a single-cycle multiplier.
\item Demonstration of an efficient MVM, achieving 20 GHz throughput and 200 ps latency.
\end{itemize}

\section{Computational Components}
\subsection{Bistable Vortex Memory}
The BVM cell \cite{Karamuftuoglu_BVM_2025} operates with four signals: word line (WL), bit line (BL), sense enable (SE), and sense line (SL). These signals facilitate write and read operations, ensuring efficient data handling in the superconducting memory structure. WL and BL currents are dedicated to writing data, while SE and SL are involved in the readout process. Unlike conventional superconducting memory, BVM eliminates the need for bulky transformers, reducing circuit complexity while maintaining high efficiency.

\begin{figure}[!t]
    \centering
    \includegraphics[width=0.98\linewidth]{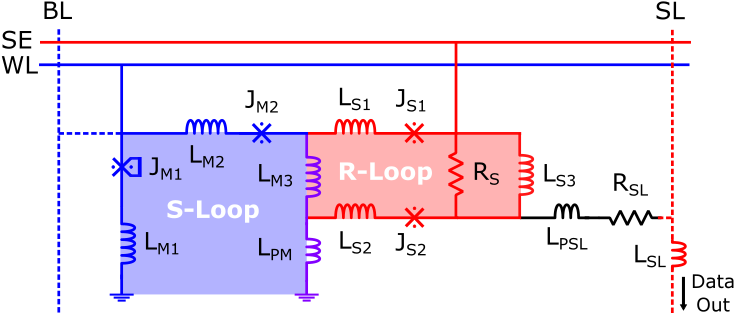}
    \caption{BVM cell model with I/O signals.}
    \label{fig:bvmCell}
\end{figure}

BVM cell consists of two superconducting loops: the Storage Loop (S-Loop) and the Readout Loop (R-Loop), each incorporating JJs and inductive components. The WL and BL currents determine the S-Loop's state by controlling the vortex circulation, representing binary 0 and 1 data as counterclockwise or clockwise circulations, respectively. During read operation, the SE signal injects additional current into the R-loop, allowing its JJs to switch according to the S-loop state. The SL captures the output current, enabling accurate data retrieval. The BVM and its signal configuration are shown in Fig. \ref{fig:bvmCell}, with color- and dash-coded lines provided in subsequent figures for clarity.

The circuit operates with zero static power dissipation as no continuous biasing or offset currents are required to maintain stored information. Consequently, BVM offers a significant advantage in its non-volatile nature, retaining data without needing a constant power supply. These properties make BVM a highly efficient and scalable memory solution for superconducting computing. We simulate BVM memory using the SPICE-based Josephson simulator (JoSIM) \cite{delportJoSIM} and model its behavior accurately. The design and properties are detailed in \cite{Karamuftuoglu_BVM_2025}.

\subsection{Quantizer Buffer}
The BVM cell generates an output current, which is input to the Quantizer Buffer (QB) cell. The QB cell acts as a thresholding element, determining whether the received current is sufficient to generate SFQ pulses \cite{razmkhah2023hybrid}. However, rather than developing a fixed number of pulses, the QB cell produces a variable number of SFQ pulses depending on the amplitude of the input current from the BVM cells. When multiple BVM cells are read simultaneously, the total current accumulates on the SL, and the resulting output amplitude scales with the number of BVM cells storing data 1.

\begin{figure}[!htbp]
\centering
\begin{subfigure}{1\linewidth}
    \centering
    \includegraphics[width=0.50\linewidth]{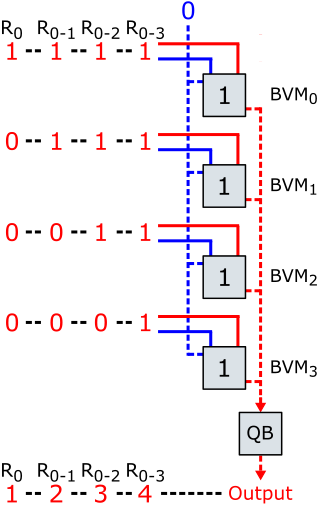}
    \caption{Testbench for the BVM array with a readout QB cell. The column here has four rows. During the read operation, the active rows are marked with a value of 1, while all others are set to 0. The QB output can range from 0 to 4 pulses, depending on how many BVM cells are accessed.}
    \label{fig:bvm4X1QB_tb}
\end{subfigure}
\hfill
\vspace{0.5mm}
\begin{subfigure}{1\linewidth}
    \centering
    \includegraphics[width=0.85\linewidth]{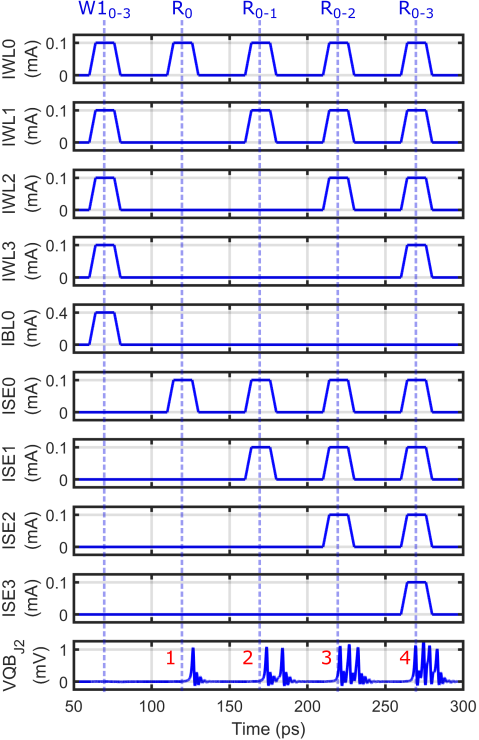}
    \caption{\textcolor{black}{Simulation result for the readout QB and BVM cells.}}
    \label{fig:bvm4X1QB}
\end{subfigure}
\caption{Evaluation of a 4-bit column BVM array with an integrated readout QB. The operations $W1_{0-3}$ and $R_{0-3}$ denote the write-1 and read processes for rows 0 through 3, respectively. In this setup, data value one is programmed into all BVM cells. When a BVM cell stores the value 1, the QB cell generates an output pulse. The number of observed pulses at the QB output varies according to the number of rows being read simultaneously.}
\label{fig:bvm4X1QBall}
\end{figure}

To better explain the interface step of the multiplier, we focus on a simple 4$\times$1 bit BVM array with a readout QB cell, configured to match the number of rows in the multiplier circuit (cf. Fig. \ref{fig:bvm4X1QBall}.) All BVM cells are programmed to store data 1 in this test setup. The current needed to generate an SFQ pulse from the QB circuit is adjusted to the output level of a single BVM cell. As a result, the number of pulses at the QB output correlates with the number of ones stored in the simultaneously accessed rows. The simulation begins with a read operation of a single row and progresses until all rows have been accessed. This approach ensures the accuracy of the intermediate step of the multiplier by generating multiple SFQ pulses that correspond to the number of BVM cells accessed, each storing a data value of 1.

\subsection{T1 Cell}
\begin{figure}[!t]
\centering
\begin{subfigure}{1\linewidth}
    \centering
    \includegraphics[width=0.95\linewidth]{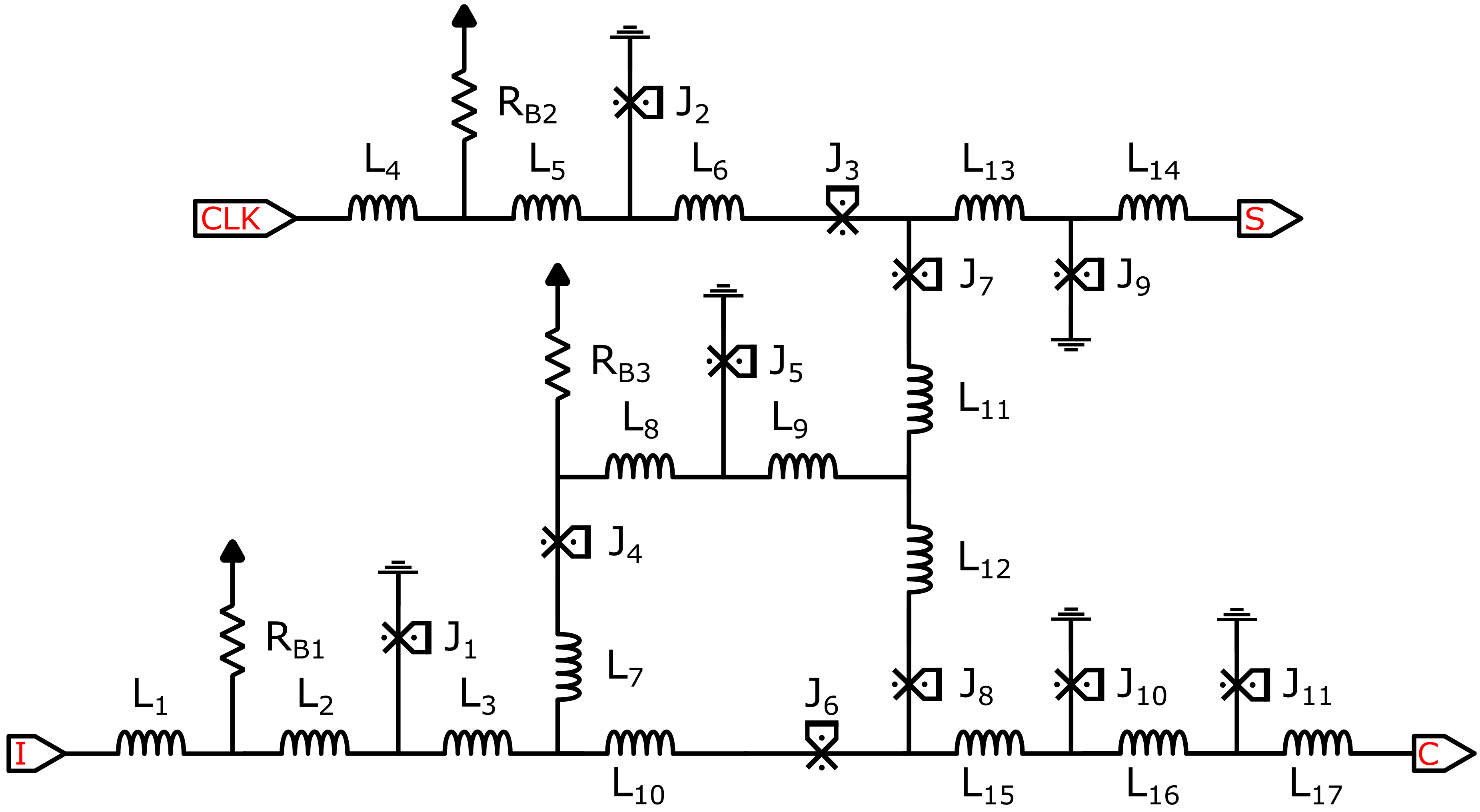}
    \caption{Schematic of T1 cell. ($L_{1}$ = 0.213pH, $L_{2}$ = 1.6pH, $L_{3}$ = 2.028pH, $L_{4}$ = 0.153pH, $L_{5}$ = 0.6pH, $L_{6}$ = 2.337pH, $L_{7}$ = 1.219pH, $L_{8}$ = 1.383pH, $L_{9}$ = 5.366pH, $L_{10}$ = 0.905pH, $L_{11}$ = 0.957pH, $L_{12}$ = 1.219pH, $L_{13}$ = 1.009pH, $L_{14}$ = 4.6pH, $L_{15}$ = 1.297pH, $L_{16}$ = 4.644pH, $L_{17}$ = 2pH,  $R_{B1}$ = $R_{B2}$ = 6.8 $\Omega$, $R_{B3}$ = 16 $\Omega$, $J_{1}$ = $J_{2}$ = $350\mu A$, $J_{3}$ = $180\mu A$, $J_{4}$ = $80.3\mu A$, $J_{5}$ = $77.9\mu A$, $J_{6}$ = $105.1\mu A$, $J_{7}$ = $J_{8}$ = $J_{9}$ = $100\mu A$, $J_{11}$ = $86.9\mu A$, $J_{11}$ = $150\mu A$)}
    \label{fig:bvmT1sch}
\end{subfigure}
\hfill
\vspace{0.5mm}
\begin{subfigure}{1\linewidth}
    \centering
    \includegraphics[width=0.70\linewidth]{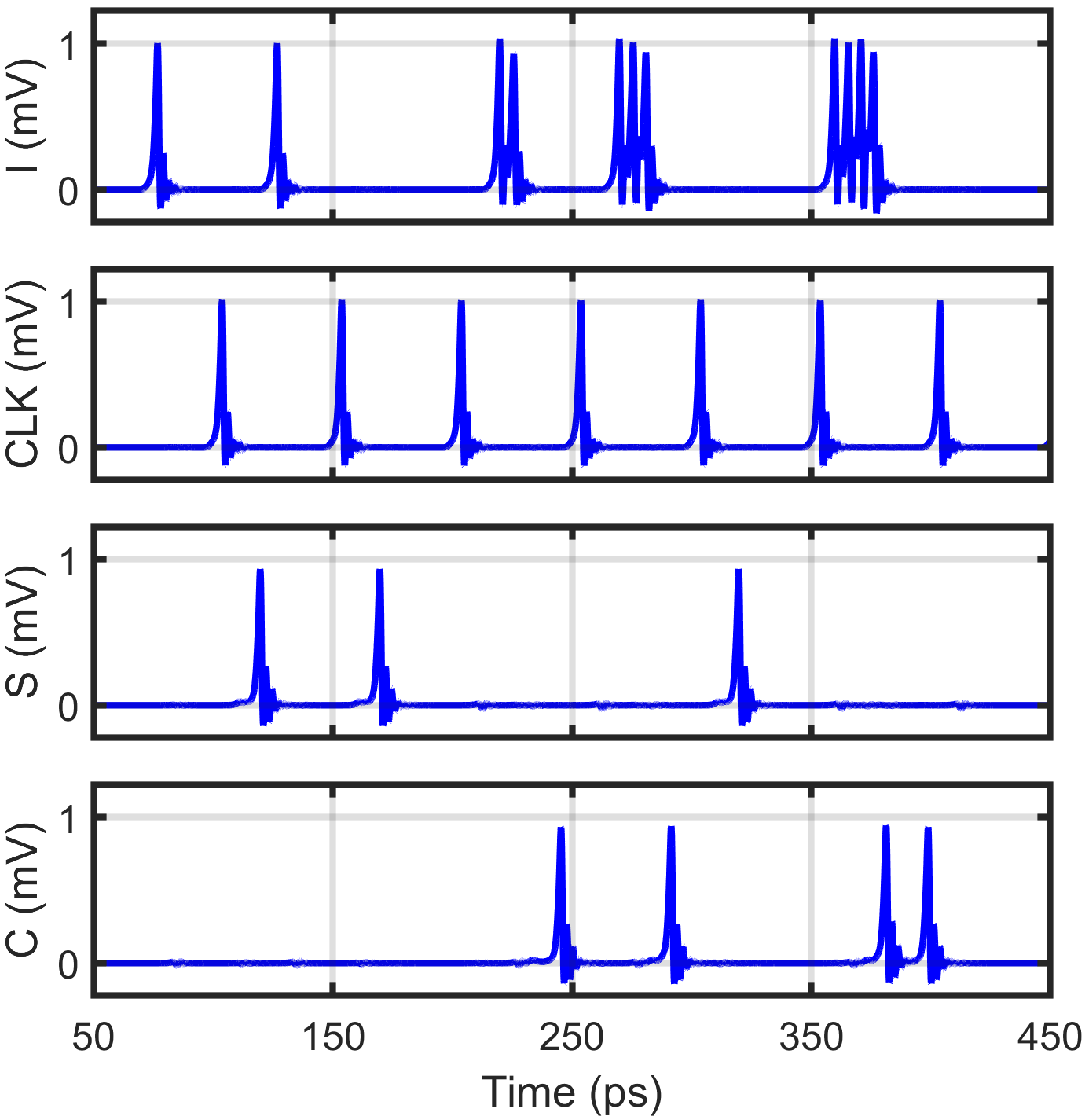}
    \caption{Simulation result. In the plots, pulses are observed from the JJs in Josephson transmission lines (JTLs) \cite{likharev1991a} connected to the T1 cell.}
    \label{fig:bvmT1sim}
\end{subfigure}
\caption{Validating functionality of the T1 cell. Input, clock, sum, and carry-out signals are denoted as $I$, $CLK$, $S$, and $C$, respectively.}
\label{fig:bvmT1all}
\end{figure}

The T1 (adder) cell is an SFQ circuit with synchronous sum and asynchronous carry output \cite{razmkhahBook, bairamkulovT1}. An SFQ pulse appears on the carry output when two or more input pulses are present, while an SFQ pulse is generated on the sum output when an odd number of input pulses are present. The related circuit with a simulation showing the generation of asynchronous carry output and synchronous sum output is provided in Fig. \ref{fig:bvmT1all}.

At the beginning of the simulation (see Fig. \ref{fig:bvmT1sim}), an SFQ pulse is applied to the T1 input, and an output pulse is generated upon receiving the CLK signal. In subsequent steps, multiple pulses are introduced in a short time frame. A carry output is generated asynchronously for every two input pulses. An SFQ pulse is observed at the sum output when the pulse count is odd. Due to the asynchronous nature of carry-out generation, multiple carry-out pulses can be observed before the arrival of the CLK signal. This behavior ensures efficient and accurate handling of pulse sequences for the BVM-based multiplier design.

\section{Multiplier design with BVM crossbar arrays}
\begin{figure}[!t]
    \centering
    \includegraphics[width=1\linewidth]{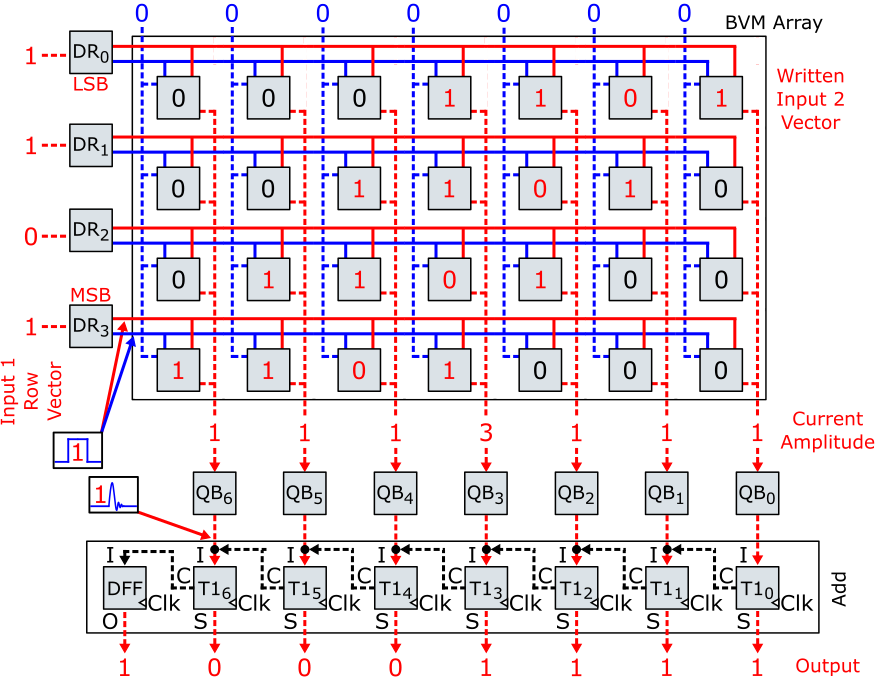}
    \caption{A 4$\times$4 bit multiplier circuit using the BVM array (preload configuration) and a carry-shifting circuit. The multiplicand (11$_d$ = 1011$_b$) is the input for the row control signal, while the multiplier (13$_d$ = 1101$_b$) is stored in the BVM array. After receiving a clock signal, the resultant product (143$_d$ = 10001111$_b$) is observed at the output. Here, the T1 cell performs binary addition with a synchronous sum and asynchronous carry output. Input, carry out, sum, and clock signals of the T1 cell correspond to $I$, $C$, $S$, and $CLK$, respectively.}
    \label{fig:bvmMult}
\end{figure}

The BVM array may be used to multiply two multi-bit numbers. As an example, we demonstrate a 4-bit multiplication. The first number (multiplicand) is written in the 4 rows of BVM so that it occupies the column positions $(3,2,1,0)$ in row 1, $(4,3,2,1)$ in row 2, $(5,4,3,2)$ in row 3, and $(6,5,4,3)$ in row 4, respectively. The second number (multiplier) is fed into the read control inputs (SE) of rows 1 through 4, initiating the read operation and enabling column-wise addition. The QB cells placed on each column perform the analog-to-digital conversion and generate one or more SFQ pulses depending on the column's output current amplitude. The outputs of QBs result from the unitary multiplication of the two numbers fed into a single-cycle carry-shifting circuit. The design is shown in Fig. \ref{fig:bvmMult}.

As seen in Fig. \ref{fig:bvmMult}, the 4$\times$4 bit multiplier circuit is configured to have four rows and seven columns, with the multiplicand and multiplier values set as 11$_d$ (1011$_b$) and 13$_d$ (1101$_b$), respectively. The multiplier number is written into the BVM array, with redundant cells storing data 0. Among the columns, the input of QB$_3$ receives a signal with triple the amplitude of a unit output current, generating three SFQ pulses. At the same time, the rest of the QBs generate one SFQ pulse each. These pulses are then directed to T1 cells, where internal carry propagation occurs. Once all T1 cells receive their respective clock signals, the final multiplication result 143$_d$ (10001111$_b$) is produced at the output.

\begin{figure}[!t]
\centering
\begin{subfigure}{1\linewidth}
    \centering
    \includegraphics[width=0.9\linewidth]{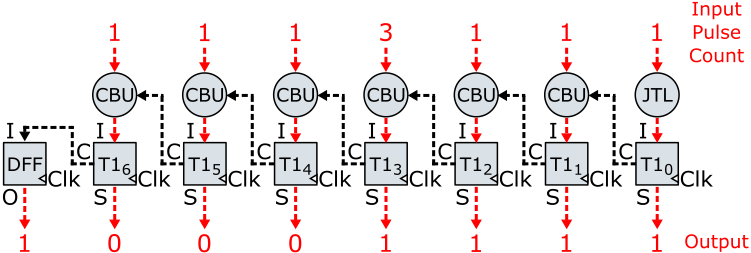}
    \caption{Testbench for the adder part of the BVM-based multiplier.}
    \label{fig:bvmAdd_tb}
\end{subfigure}
\hfill
\vspace{0.5mm}
\begin{subfigure}{1\linewidth}
    \centering
    \includegraphics[width=0.85\linewidth]{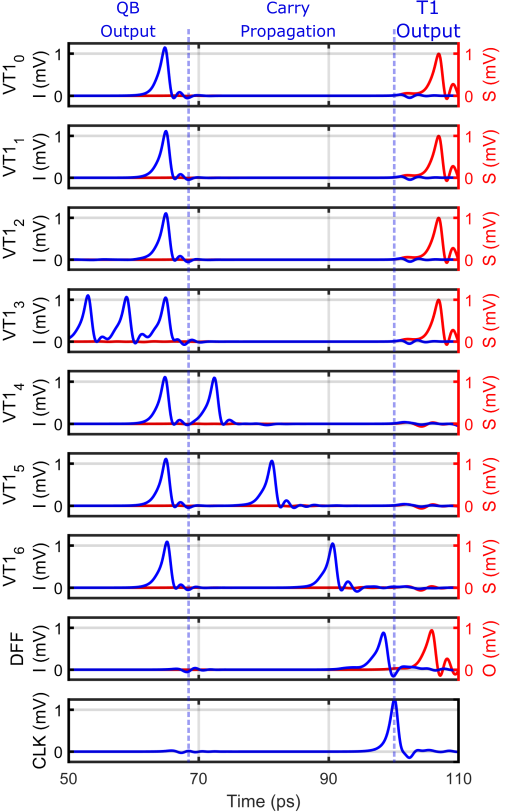}
    \caption{Simulation result for the multiplication of 11$_d$ $\times$ 13$_d$ following the QB outputs shown in Fig. \ref{fig:bvmMult}. The pulses for both the T1 input and output are detected through $J_1$ within T1 and the JJs in the JTLs connected to T1.}
    \label{fig:bvmAddsim}
\end{subfigure}
\caption{Demonstration of the adder used in the multiplier design. The QB outputs can be multiple SFQ pulses on each column. Therefore, a carry-shifting operation is required in the last stage of the multiplier. For the T1, $I$ and $S$ correspond to the input and sum signals, and for the DFF, $I$ and $O$ represent the input and output signals, respectively.}
\label{fig:bvmAddall}
\end{figure}

When the input pulses arrive from QBs to T1s, the carry bits are propagated asynchronously from the least significant to the most significant bits. Unlike the standard carry propagation methods employed in conventional multiplier structures, this propagation is only performed once, from the least significant to the most significant bits, at the end of the process using a ripple carry T1-based adder. The output is then generated with the arrival of a single clock pulse. To increase the fan-in of the T1 cell, we add a confluence buffer (CBU) to the structure while matching the delay of the first column with the Josephson transmission line (JTL) \cite{likharev1991a}. An example of this configuration is shown in Fig. \ref{fig:bvmAddall}.

For the example of multiplication of 11$_d$ and 13$_d$ as in Fig.~\ref{fig:bvmAddsim}, all T1s receive an SFQ pulse except T1$_{3}$, having three SFQ pulses that appear before the 70~ps time point. Since the amplitude of the BVM signal on the QB$_{3}$'s column is larger than the rest of the columns, this QB cell starts generating SFQ pulses earlier than the rest. In this example, the carry propagation occurs between the 70~ps and 100~ps time points. The resultant product (143$_d$ = 10001111$_b$) is observed when a clock pulse (Clk) is applied to the cells.

To compare the multiplier designed using the BVM array with one created using conventional RSFQ logic, we have synthesized a 4-bit multiplier using the ColdFlux library and tools \cite{Coldflux2023}. The conventional multiplier circuit uses 13,117 JJs, with an area of 4.14~mm$^2$ and a logic gate depth of 12. The circuit can operate at 44.4 GHz clock frequency, producing 270~ps latency and consuming about 1.93~mW of static and 0.05~mW of dynamic power on average. The proposed design with the BVM design has about ~550 JJs, with an area of 0.236~mm$^2$. The initialization for the multiplier value takes five clock cycles and generates the output with one clock cycle for every new multiplicand, operating at 20~GHz and resulting in 50~ps latency. The estimated static power consumption for this design is 0.145~mW, including the peripherals, and consumes 12.23~$\mu$W of dynamic power.

\subsection{Optimized BVM Array Implementation for Multiplication}
The original BVM array can be utilized for memory operations and multiplier computations. However, this dual functionality comes at the cost of increased area usage and additional initialization cycles, which impact overall performance. To address this, we have restructured the SLs to eliminate unused memory cells in the top-left and bottom-right regions of the array. This modification eliminates 12 BVM cells and three BLs from the design. Connecting the SLs diagonally ensures accurate current accumulation, optimizes circuit area, and improves efficiency.

In the initial design, one operand is provided as a row input, while the other is stored in the memory array. However, repeatedly writing the same operand across multiple memory rows introduces inefficiencies. BLs can supply the second operand to address this, eliminating redundant memory writes during initialization. This approach requires a uniform preloading of the memory array with a value of 1. The memory cells storing 1 act as an AND mask, generating an output only when both row and column inputs are 1. This modification enables the generation of results after a single initialization cycle, significantly enhancing computational efficiency.

\begin{figure}[!t]
    \centering
    \includegraphics[width=0.92\linewidth]{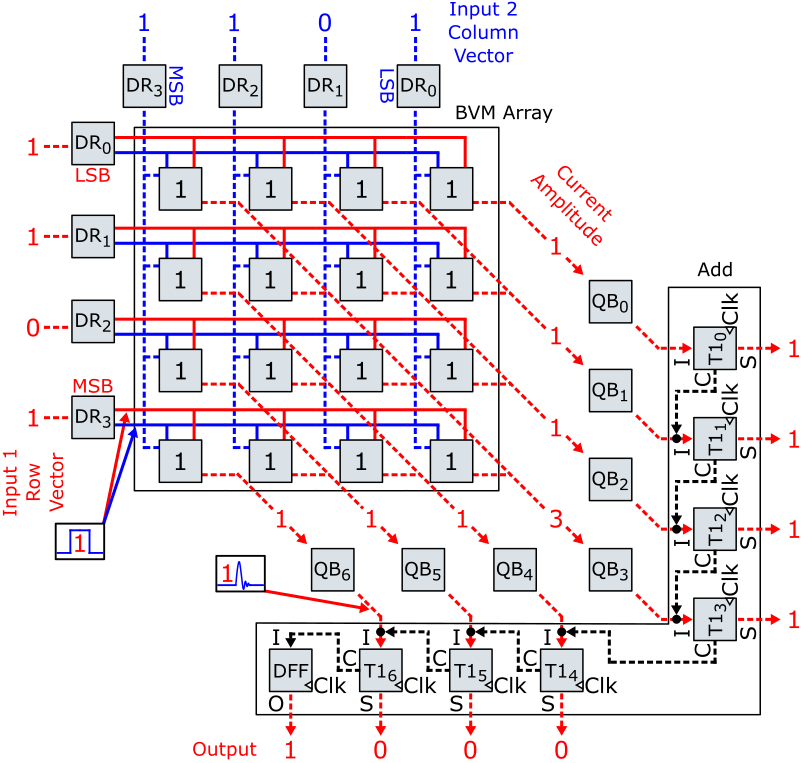}
    \caption{Modified BVM array (direct-input configuration) for 4$\times$4 multiplication. SLs are connected diagonally to enable accumulation with fewer BVM cells. The read operation is carried out using SE for rows and BL for columns. The data values and pin names are consistent with those in Fig. \ref{fig:bvmMult}.}
    \label{fig:bvmMultImproved}
\end{figure}

The first approach requires reinitialization whenever the multiplier operand changes, increasing latency. However, storing only 1s in memory mitigates the need for frequent initialization, further improving performance. The input scheme must be adjusted for the improved approach, as WL and BL intersect at the same node in a BVM cell. Rather than using only row-based read operations with WL and SE, inputs should be provided through SE for rows and BL for columns to ensure a proper multiplication function. The modified design implementing this optimization is illustrated in Fig. \ref{fig:bvmMultImproved}.

\subsection{Multiply-Accumulate Implementation}
To enable MAC operation, existing T1 cells used for multiplication can be extended by adding additional T1 cells along with D flip-flops (DFF). Although T1 cells handle final carry propagation during multiplication, simply increasing the number of T1 cells at the output is not sufficient, as this would introduce extra carry delays and reduce performance. Instead, additional T1 cells must be strategically placed with DFFs to properly manage carry propagation while maintaining the original clock frequency.

\begin{figure}[!t]
    \centering
    \includegraphics[width=0.86\linewidth]{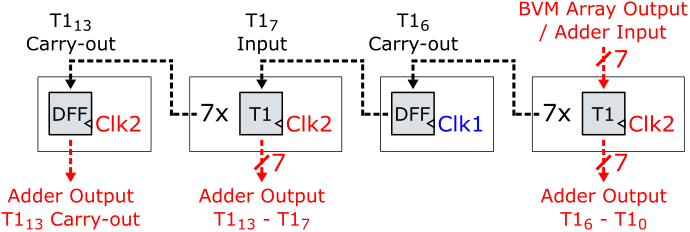}
    \caption{Extended adder design in multiplier to implement MAC operation.}
    \label{fig:bvmMACImproved}
\end{figure}

The DFFs retain the carry information across accumulation cycles. For example, to generate a 15-bit accumulation result, the output is divided after the lower 7 bits, requiring an additional group of T1 cells. As shown in Fig.~\ref{fig:bvmMACImproved}, the lower part of the result is processed by T1 cells $T1_{6-0}$, while the upper part is handled by $T1_{13-7}$. The carry generated by $T1_6$ is passed into the upper section, allowing the full 15-bit accumulated result to be correctly calculated.

To ensure proper timing synchronization during carry propagation, the clock for the intermediate DFF must be distinct from the others. In this configuration, the intermediate DFF receives a dedicated clock signal (Clk1) for each new multiplication result accumulated, while the remaining cells operate with Clk2 to generate the final accumulation output. Additional DFFs and seven-T1 configurations can be incorporated for higher-bit accumulation as needed. However, increasing the number of intermediate DFFs introduces additional clock cycles for carry propagation, with the required delay depending on the total count of intermediate DFF stages.

\section{Matrix-Vector Multiplication}

MVM is fundamental in various computing applications, including neural networks and signal processing. It involves computing the dot product of each row of the matrix with the input vector, producing an output vector where each element represents a weighted sum of the corresponding row elements. Given a matrix \( W \) of size \( n \times m \) and a vector \( x \) of size \( n \times 1 \), the resulting output vector \( y \) of size \( m \times 1 \) is calculated as follows:

\begin{equation}
y_i = \sum_{j=1}^{n} {W^T}_{ij} x_j
\label{eq:mvmFormula}
\end{equation}

where \( i \) and \( j \) denote the row and column indices, respectively. For demonstration, we consider a \( 4 \times 4 \) example:

\begin{equation}
\begin{bmatrix}
    W_{11} & W_{12} & W_{13} & W_{14} \\
    W_{21} & W_{22} & W_{23} & W_{24} \\
    W_{31} & W_{32} & W_{33} & W_{34} \\
    W_{41} & W_{42} & W_{43} & W_{44} 
\end{bmatrix}^{\mathrm{T}}
\begin{bmatrix}
    x_1 \\
    x_2 \\
    x_3 \\
    x_4
\end{bmatrix}
=
\begin{bmatrix}
    y_1 \\
    y_2 \\
    y_3 \\
    y_4
\end{bmatrix}
\label{eq:mvm4x4}
\end{equation}

To efficiently realize MVM, we employ BVM-based multipliers, which are designed for high-speed and low-power operations. The calculation based on the design in Fig. \ref{fig:bvmMult} follows a structure similar to memristive crossbar arrays, where each matrix element \( W_{ij} \) is stored in the multiplier array based on BVM, while the input vector \( x \) is provided as row inputs. Regardless of whether the original design in Fig. \ref{fig:bvmMult} or the modified version in Fig. \ref{fig:bvmMultImproved} is used, the overall architecture remains unchanged. The only difference lies in the initialization phase, where the values stored in memory are preloaded or provided dynamically as input. In the modified version, the required values are provided directly during computation, rather than writing one of the operands into memory beforehand.

\begin{figure}[!t]
    \centering
    \includegraphics[width=0.92\linewidth]{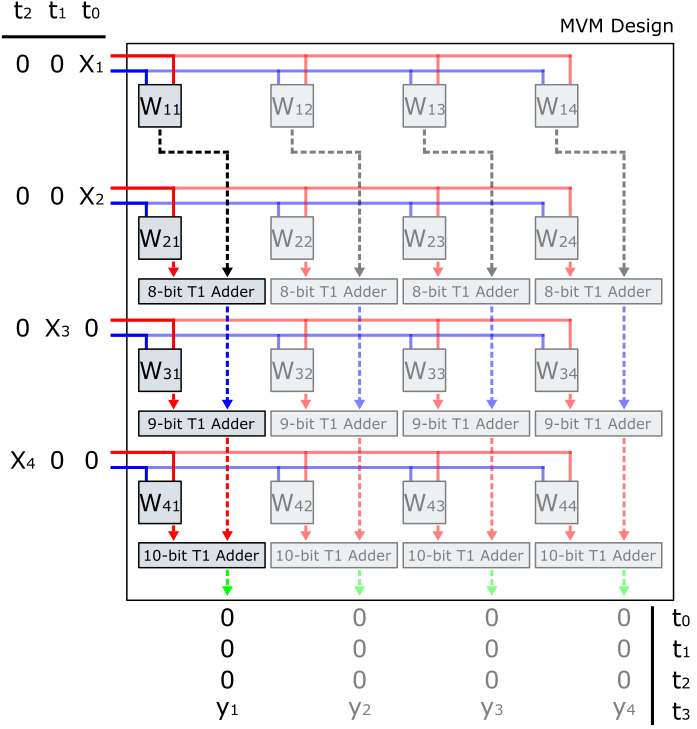}
    \caption{MVM structure utilizing the BVM-based multiplier. In this preload configuration, $W$ values are stored in BVM arrays, whereas in the non-preload (direct-input) version, they are expected to be provided dynamically through column inputs. For clarity, only the multipliers and data flow are shown, while DFFs and other components are omitted.}
    \label{fig:mvmStructure}
\end{figure}

The multiplication process occurs in parallel due to the tiled BVM multipliers, and T1 cells handle the accumulation of partial sums for the final computation. Note that the inputs are deliberately delayed to match their timing on the adders. As shown in Fig.~\ref{fig:mvmStructure}, the proposed architecture integrates BVM multipliers into a tiled structure, enabling scalable and energy-efficient MVM computation.

To demonstrate this approach, we consider a \( 4 \times 4 \) matrix \( W \) and a 4-element input vector \( x \) as an example. The matrix elements are preloaded into the BVM array, while \( x \) is provided sequentially as input. The tiled multipliers execute the multiplications simultaneously, and the products are accumulated row-wise to generate the final output vector \( y \). In Eq. \ref{eq:mvmExample}, the highlighted values indicate the input elements used to perform the computation and the corresponding result. The first multiplication corresponds to the operation already demonstrated in Fig.~\ref{fig:bvmAddall}.

\begin{equation}
\begin{bmatrix}
    \textcolor{blue}{13} & \textcolor{blue}{7} & \textcolor{blue}{12} & \textcolor{blue}{15} \\
    4 & 0 & 5 & 13 \\
    2 & 1 & 12 & 11 \\
    14 & 15 & 2 & 6
\end{bmatrix}
\begin{bmatrix}
    \textcolor{blue}{11} \\
    \textcolor{blue}{10} \\
    \textcolor{blue}{9} \\
    \textcolor{blue}{15}
\end{bmatrix}
=
\begin{bmatrix}
    \textcolor{blue}{546} \\
    284 \\
    305 \\
    412
\end{bmatrix}
\label{eq:mvmExample}
\end{equation}

\begin{figure}[!t]
    \centering
    \includegraphics[width=0.88\linewidth]{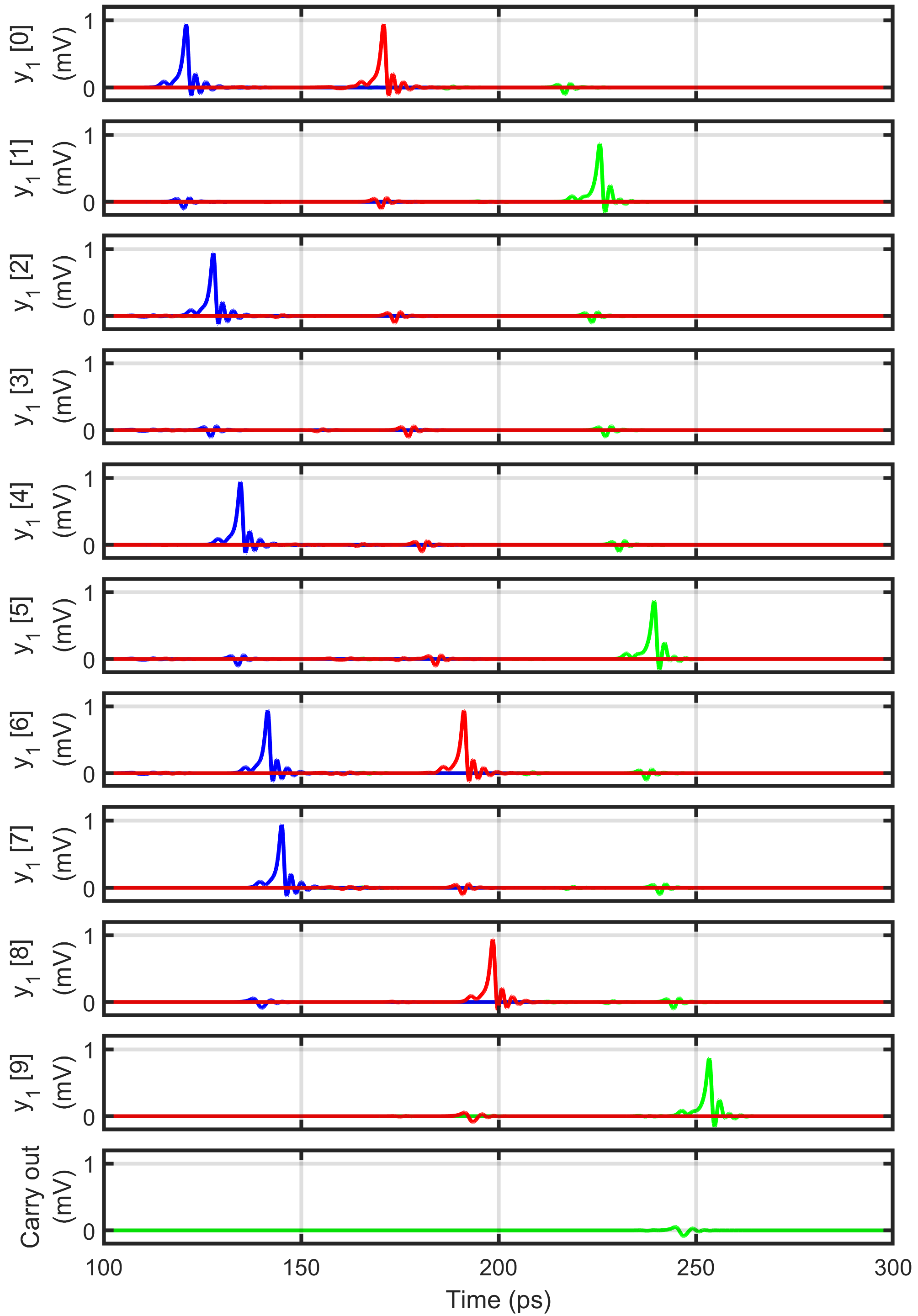}
    \caption{Simulation of data accumulation on MVM, showing the computation of \( y_1 \) as derived from Eq. \ref{eq:mvmExample}. At \( t_0 \), the initial multiplications are completed by the multipliers storing \( W_{11} \) and \( W_{21} \). The resulting sum, 213$_d$, is computed by the 8-bit T1 adder at \( t_1 \) and is highlighted in blue. At \( t_2 \), the design generates a 9-bit intermediate result of 321$_d$, marked in red. Finally, at \( t_3 \), the accumulation completes with a 10-bit result of 546$_d$, highlighted in green.}
    \label{fig:mvmSimulation}
\end{figure}

After four clock cycles, the proposed design generates the output vector \( y \) operating at 20 GHz. Figure \ref{fig:mvmSimulation} shows the simulation results for the values highlighted in Equation \ref{eq:mvmExample}. At $t_0$, the multipliers with $W_{11}$ and $W_{21}$ perform the initial multiplications. The summation result (213$_d$ = 11010101$_b$) of their outputs, highlighted in blue, is observed from the 8-bit T1 adder at $t_1$. Consequently, the design generates the 9-bit result of 321$_d$ = 101000001$_b$ in red at $t_2$ and the 10-bit result of 546$_d$ = 1000100010$_b$ in green at $t_3$. 

The simulation evaluates the performance of the proposed MVM design. The results confirm the expected matrix-vector multiplication behavior, in particular verifying the multiplication of the first matrix row with the input vector. Implementing first-row multiplication requires about 2708 JJs and results in an estimated static power consumption of 0.728 mW. Furthermore, the design can be extended to realize matrix-matrix multiplication with the systolic array approach, enabling efficient and scalable computation for more complex workloads. These findings demonstrate the feasibility of the proposed architecture for high-speed, large-scale superconducting computing applications.

\section{Conclusion}
This work introduces an efficient and scalable matrix-vector multiplication (MVM) architecture, leveraging Bistable Vortex Memory (BVM) based multipliers, thereby extending the application of the BVM concept previously introduced for data storage. The central contribution is demonstrating how the BVM's capability for current summation can be effectively utilized within a multiplier design by integrating BVM arrays with Quantizer Buffer (QB) and T1 adder cells. This integration creates a high-speed, low-power computational unit capable of performing multiplication and accumulation directly within the memory array. The proposed MVM design utilizes a tiled multiplier structure that facilitates parallel computation. A significant innovation presented is the development of an optimized BVM array implementation specifically for multiplication, featuring a restructured Sense Line (SL) configuration with diagonal connections and a revised input method, which results in a smaller area and improved efficiency compared to the original BVM memory array design. As demonstrated in this work, BVM arrays, when configured as multipliers, can perform column addition/accumulation in a single cycle, making this design an efficient in-memory and near-memory computational unit suitable for matrix-vector multiplication, neural networks, and combinatorial logic implementations. The designs are validated through simulations, demonstrating correct computation and high-speed performance at 20 GHz. By presenting BVM's successful application in computational structures like multipliers and MAC units, this work emphasizes its versatility beyond purely memory functions, offering an efficient alternative to conventional digital implementations for these critical tasks.

\ackn This work has been funded by the National Science Foundation (NSF) under the project Expedition: (Design and Integration of Superconducting Computation for Ventures beyond Exascale Realization) with grant number 2124453. The authors thank Arda Caliskan (USC) for providing the T1 cell.

\section*{References}
\bibliographystyle{iopart-num}
\bibliography{references}

\providecommand{\newblock}{}
\begin{thebibliography}{10}
\expandafter\ifx\csname url\endcsname\relax
  \def\url#1{{\tt #1}}\fi
\expandafter\ifx\csname urlprefix\endcsname\relax\def\urlprefix{URL }\fi
\providecommand{\eprint}[2][]{\url{#2}}

\bibitem{holmes2021cryogenic}
Holmes D~S 2021 Cryogenic electronics and quantum information processing {\em 2021 IEEE International Roadmap for Devices and Systems Outbriefs\/} (IEEE) pp 1--93

\bibitem{razmkhahBook}
{Razmkhah, Sasan and Febvre, Pascal} 2023 {Superconducting Quantum Electronics} {\em {Beyond-CMOS}\/} (ISTE \& WILEY) chap~8, pp 295--391 ISBN 9781394228713

\bibitem{likharev1991a}
Likharev K~K and Semenov V~K 1991 {\em IEEE transactions on applied superconductivity\/} {\bf 1} 3--28

\bibitem{takeuchiAQFP2013}
Takeuchi N, Ozawa D, Yamanashi Y and Yoshikawa N 2013 {\em Superconductor Science and Technology\/} {\bf 26} 035010

\bibitem{razmkhah2024challenges}
Razmkhah S, Aviles R~S, Li M, Gupta S, Beerel P~A and Pedram M 2024 Challenges and unexplored frontiers in electronic design automation for superconducting digital logic {\em 2024 Design, Automation \& Test in Europe Conference \& Exhibition (DATE)\/} (IEEE) pp 1--6

\bibitem{Haolin_Multiplier_2021}
Cong H, Li M and Pedram M 2021 {\em IEEE Transactions on Applied Superconductivity\/} {\bf 31} 1--10

\bibitem{Nagaoka_Multiplier_2021}
Nagaoka I, Ishida K, Tanaka M, Sano K, Yamashita T, Ono T, Inoue K and Fujimaki A 2021 {\em IEEE Transactions on Applied Superconductivity\/} {\bf 31} 1--5

\bibitem{Yamanashi_Multiplier_2024}
Yamanashi Y, Okumura H and Yoshikawa N 2024 {\em Superconductor Science and Technology\/} {\bf 37} 115024

\bibitem{Coldflux2023}
Fourie C~J, Jackman K, Delport J, Schindler L, Hall T, Febvre P, Iwanikow L, Chen O, Ayala C~L, Yoshikawa N {\em et~al.\/} 2023 {\em IEEE Transactions on Applied Superconductivity\/} {\bf 33} 1--26

\bibitem{McCanny_MVM_1983}
McCanny J and McWhirter J 1983 {\em IEE Proceedings G (Electronic Circuits and Systems)\/} {\bf 130}(4) 125--130

\bibitem{Alam_Superconducting_MVM_2023}
Alam S, Hutchins J, Hossain M~S, Ni K, Narayanan V and Aziz A 2023 Cryogenic in-memory matrix-vector multiplication using ferroelectric superconducting quantum interference device (fe-squid) {\em 2023 60th ACM/IEEE Design Automation Conference (DAC)\/} pp 1--6

\bibitem{zolfagharinejad2024brain}
Zolfagharinejad M, Alegre-Ibarra U, Chen T, Kinge S and van~der Wiel W~G 2024 {\em The European Physical Journal B\/} {\bf 97} 70

\bibitem{bao2022toward}
Bao H, Zhou H, Li J, Pei H, Tian J, Yang L, Ren S, Tong S, Li Y, He Y {\em et~al.\/} 2022 {\em Frontiers of Optoelectronics\/} {\bf 15} 23

\bibitem{siegl2016data}
Siegl P, Buchty R and Berekovic M 2016 Data-centric computing frontiers: A survey on processing-in-memory {\em Proceedings of the Second International Symposium on Memory Systems\/} pp 295--308

\bibitem{seshadri2017ambit}
Seshadri V, Lee D, Mullins T, Hassan H, Boroumand A, Kim J, Kozuch M~A, Mutlu O, Gibbons P~B and Mowry T~C 2017 Ambit: In-memory accelerator for bulk bitwise operations using commodity dram technology {\em Proceedings of the 50th Annual IEEE/ACM International Symposium on Microarchitecture\/} pp 273--287

\bibitem{chua_memristor-missing_1971}
Chua L 1971 {\em IEEE Transactions on Circuit Theory\/} {\bf 18} 507--519 ISSN 0018-9324 \urlprefix\url{http://ieeexplore.ieee.org/document/1083337/}

\bibitem{strukov2008missing}
Strukov D~B, Snider G~S, Stewart D~R and Williams R~S 2008 {\em nature\/} {\bf 453} 80--83

\bibitem{williams2008we}
Williams R~S 2008 {\em IEEE spectrum\/} {\bf 45} 28--35

\bibitem{sun2019understanding}
Sun W, Gao B, Chi M, Xia Q, Yang J~J, Qian H and Wu H 2019 {\em Nature communications\/} {\bf 10} 3453

\bibitem{zidan2018future}
Zidan M~A, Strachan J~P and Lu W~D 2018 {\em Nature electronics\/} {\bf 1} 22--29

\bibitem{lee2018demand}
Lee J and Lu W~D 2018 {\em Advanced Materials\/} {\bf 30} 1702770

\bibitem{yao2020fully}
Yao P, Wu H, Gao B, Tang J, Zhang Q, Zhang W, Yang J~J and Qian H 2020 {\em Nature\/} {\bf 577} 641--646

\bibitem{han2020electric}
Han Y, Nickle C, Zhang Z, Astier H~P, Duffin T~J, Qi D, Wang Z, Del~Barco E, Thompson D and Nijhuis C~A 2020 {\em Nature materials\/} {\bf 19} 843--848

\bibitem{jung2022crossbar}
Jung S, Lee H, Myung S, Kim H, Yoon S~K, Kwon S~W, Ju Y, Kim M, Yi W, Han S {\em et~al.\/} 2022 {\em Nature\/} {\bf 601} 211--216

\bibitem{Karamuftuoglu_BVM_2025}
Karamuftuoglu M, Ucpinar B, Razmkhah S and Pedram M 2024 {\em Superconductor Science and Technology\/} {\bf 38} 015020

\bibitem{bairamkulovT1}
Bairamkulov R, Yu M and De~Micheli G 2024 {Unleashing the Power of T1-cells in SFQ Arithmetic Circuits} {\em Proceedings of the 61st ACM/IEEE Design Automation Conference\/} pp 1--6

\bibitem{delportJoSIM}
Delport J~A, Jackman K, Le~Roux P and Fourie C~J 2019 {\em IEEE Transactions on Applied Superconductivity\/} {\bf 29} 1--5

\bibitem{razmkhah2023hybrid}
Razmkhah S, Karamuftuoglu M~A and Bozbey A 2024 {\em Superconductor Science and Technology\/} {\bf 37} 065011

\end{thebibliography}
\end{document}